\documentclass[11pt]{article}
\usepackage{epsfig}
\usepackage{amscd}
\usepackage{amssymb}
\usepackage{amsfonts}
\usepackage{amsmath}
\usepackage{mathrsfs}
\newcommand{\be}{\begin{equation}}
\newcommand{\ee}{\end{equation}}
\newcommand{\ba}{\begin{eqnarray}}
\newcommand{\ea}{\end{eqnarray}}
\newcommand{\lpb}{\{\!\!\{}
\newcommand{\rpb}{\}\!\!\}}
\newcommand{\half}{\tfrac{1}{2}}
\usepackage{graphics,epsfig,psfrag}

\begin{document}

\makeatletter
\@addtoreset{equation}{section}
\makeatother
\renewcommand{\theequation}{\thesection.\arabic{equation}}
\renewcommand{\thefootnote}{\alph{footnote}}

\begin{titlepage}

\baselineskip =15.5pt
\pagestyle{plain}
\setcounter{page}{0}

\vfil

\begin{center}
{\LARGE {\bf Path Integrals for (Complex) Classical and Quantum Mechanics}}
\end{center}

\vfil
\begin{center}
{\large R.J.~Rivers
}\\
\vspace {1mm}
Physics Department, Imperial College London, \\
London SW7 2AZ, UK \\
e-mail: {\it r.rivers@imperial.ac.uk}
\end{center}

\vfil

\begin{abstract}
\noindent An analysis of classical mechanics in a complex extension of phase space shows that a particle in such a space can behave in a way redolant of quantum mechanics; additional degrees of freedom permit 'tunnelling' without recourse to instantons and lead to time/energy uncertainty.
In practice, 'classical' particle trajectories with additional degrees of freedom have arisen in several different formulations of quantum mechanics. In this talk we compare the extended phase space of the closed time-path formalism with that of complex classical mechanics, to suggest that $\hbar$ has a role in our understanding of the latter. However, differences in the way that trajectories are used make a deeper comparison problematical. We conclude with some thoughts on quantisation as dimensional reduction.
\end{abstract}

\end{titlepage}
\newpage

\section{Introduction}
 It has been argued recently, in a series of papers by Carl Bender and collaborators
 \cite{bender1,bender2,bender3,bender3a}, that classical mechanics in a {\it complex} extension of phase space has some attributes of  quantum mechanics.
 With these additional dimensions particles can negotiate otherwise impenetrable classical hills, enabling them to move from one potential well to another as an alternative to quantum tunnelling.  Further, classical particles with complex energy in periodic potentials can exhibit a kind of band structure\cite{bender3a}. However, the discussion has now gone beyond the mere cataloguing of these qualitative similarities to suggesting that, for the case of 'tunnelling', for example, we are seeing behaviour that can approximate the probabilistic results of quantum mechanics {\it quantitatively} \cite{bender3}. Although the emphasis has been on mimicking $\hbar$-independent (at leading order) probability ratios, for this procedure to be sensible $\hbar$ must be implicit in the analysis, since the formalism of complex classical mechanics cannot, of itself, distinguish between {\it quantum} (action $O(\hbar)$) systems and {\it classical} systems (action $O (\hbar^0)$). In fact, should there be any relationship between complex classical mechanics and quantum mechanics, there is a strong hint as to how this must happen, since the tunnelling complies with time/energy uncertainty relations \cite{bender1}, albeit with no $\hbar$ visible.

 On the other hand, if we adopt the contrary position of extending (real) classical behaviour to  quantum mechanical behaviour, it is well-known that there are several different ways in which classical or 'quasi-classical' paths can arise in the formulation of quantum dynamics, for which the role of $\hbar$ is clear. In particular, we are interested in what we might term a Moyal path integral approach which (in co-ordinate space) reduces to the Feynman-Vernon closed time-path approach \cite{feynver} for the evolution of density matrices, familiar in the analysis of decoherence. In fact, it was thinking about the earlier use  of classical trajectories by the author and collaborators \cite{rivers,rivers2} to approximate the evolution of the quantum density matrix  that provoked this talk. In this commentary we shall contrast the quasi-classical paths of this Moyal formalism to the classical paths in complex phase-space discussed in \cite{bender3}, in each case the solutions to a constrained Hamiltonian system.

 For reasons that will rapidly become clear, we shall restrict ourselves to Hermitian Hamiltonians, even though part of the original motivation for considering complex phase space was to accommodate pseudo-Hermitian Hamiltonians which, by virtue of PT-symmetry \cite{BB,B,B2}, had real spectra. 
 In the next two sections we introduce path integrals for {\it real} classical mechanics, as developed by Gozzi over several years \cite{gozzi1a,gozzi1b,gozzi1c} and show how the Moyal path integrals for quantum mechanical systems evolve naturally from them, as originally proposed by Marinov \cite{marinov} and Gozzi \cite{gozzi2a,gozzi2b}. We show that the dynamics of the quasi-classical trajectories that provide the mean-field approximation to this quantum system has the same symplectic structure as the trajectories of complex classical mechanics shown by Smilga \cite{smilga}.

 Insofar as the two approaches have some similarities  we may hope to impute $\hbar$ behaviour to complex classical mechanics.
 On the other hand, insofar as they have differences it might be thought that, the better that Moyal paths represent quantum mechanics, to which they are an established approximation, the poorer the paths from CCM described in \cite{bender3} will do so. In fact, as we shall see, the situation is somewhat more complicated.

  Even prior to the papers of \cite{marinov} and \cite{gozzi2a,gozzi2b} there was an extensive literature on the role on classical and quasi-classical paths in quantum mechanics (e.g. see \cite{gutzwiller}) that has been developed since. Given the brevity and simplicity of our observations it is sufficient, in the context of this paper, to cite just \cite{dittrich} and the references therein for interested readers. We conclude with some thoughts on quantisation as dimensional reduction.

\section{Phase space path integrals for (real) classical mechanics:}

We restrict ourselves to a single particle, mass $m$, moving in phase space $\varphi = (x, p)$, under the Hamiltonian
\begin{equation}
 H_{cl}(\varphi) := H_{cl}(x,p) = \frac{p^2}{2m} + V(x).
 \nonumber
 \label{H}
\end{equation}
The classical solutions $\varphi_{cl}$ satisfy Hamilton's equations
\be
\dot{\varphi}^a -\omega^{ab}\partial_b H_{cl}= 0,
\label{Ham0}
\ee
where $\omega_{ab} = i\sigma_2$ and $\partial_a = \partial/\partial\varphi^a$.

Classical phase-space densities $\rho_{cl}(\varphi, t)$ evolve as
\be
\rho_{cl} (\varphi_f, t_f) = \int {\cal D}\varphi_a~K_{cl}(\varphi_f,t_f|\varphi_i,t_i)\rho_{cl}(\varphi_i,t_i)
\label{clev}
\ee
where the kernel $K_{cl}(\varphi_f,t_f|\varphi_i,t_i)$ is restricted to the classical paths of (\ref{Ham0}),
\begin{eqnarray}
 K_{cl}(\varphi_f,t_f|\varphi_i,t_i)
 &=& \int_{\varphi_i}^{\varphi_f}{\cal D}\varphi_a~\delta [{\varphi}^a -\varphi_{cl}^a]
 = \int_{\varphi_i}^{\varphi_f}{\cal D}\varphi_a~\delta [\dot{\varphi}^a -\omega^{ab}\partial_b H_{cl}]
 \label{cl1}
\end{eqnarray}
The second equality of (\ref{cl1}) is a consequence of the incompressibility of phase space.

We repeat the earlier analysis of Gozzi \cite{gozzi1a} in doubling phase space through the functional Fourier transform of the delta functional:
\be
 K_{cl}(\varphi_f,t_f|\varphi_i,t_i) = \int_{\varphi_i}^{\varphi_f} {\cal D}\varphi^a {\cal D}\Pi_a~\exp\bigg\{i\int_{t_i}^{t_f}dt~\Pi_a(\dot{\varphi}^a -\omega^{ab}\partial_b H_{cl})\bigg\}
 \label{Kcl1}
\ee
If $\langle ...\rangle$ denotes averaging with respect to the (normalised) path integral then, on time-splitting, we see that the $\Pi_a$ are conjugate to the $\varphi^a$, with equal-time commutation relations
\be
 \langle [\varphi^a,\Pi_b]\rangle = i\delta^a_b
 \label{CCR}
\ee
That is, in (\ref{Kcl1}) we have the path integral realisation of the canonical Koopman - von Neumann (KvN) Hilbert space description of classical mechanics\cite{K,vN}.
This becomes clearer if we rewrite $K_{cl}(\varphi_f,t_f|\varphi_i,t_i)$ as
 \be
 K_{cl}(\varphi_f,t_f|\varphi_i,t_i) = \int_{\varphi_i}^{\varphi_f} {\cal D}\varphi^a {\cal D}\Pi_a~\exp\bigg\{i S_{cl}[\varphi,\Pi]\bigg\}
 \label{Kcl2}
\ee
where
\be
S_{cl}[\varphi,\Pi] =\int_{t_i}^{t_f}dt~{\cal L}_{cl}(\varphi,\Pi)
\ee
with
\be
 {\cal L}_{cl}(\varphi,\Pi) = {\dot\varphi^a}\Pi_b - {\cal H}_{cl}(\varphi,\Pi).
 \label{lcl}
 \nonumber
 \ee
The Hamiltonian ${\cal H}_{cl}(\phi,\Pi)$,
  \be
 {\cal H}_{cl}(\varphi,\Pi) = \Pi_a\omega^{ab}\partial_b H_{cl}.
 \label{hcl}
 \ee
 is no more than the Liouville operator (up to a factor of $i$), as follows immediately if we adopt the $\varphi$-representation $\Pi_a = -i\partial_a$ in ${\cal H}_{cl}$.
The commutators with ${\cal H}_{cl}(\varphi,\Pi)$ (in the sense of (\ref{CCR})) determine the evolution of $(\varphi,\Pi)$, but it is more convenient to think of these solutions as just the solutions $(\varphi_{cl}, \Pi_{cl})$ to $\delta S_{cl} = 0$.

\section{Phase Space Path Integrals for Quantum Mechanics:}

To explore the role of semi-classical paths in quantum mechanics we stay as close to the classical formalism of the previous section as possible. The classical phase-space density $\rho_{cl}(\varphi, t)$ is replaced by the Wigner function $\rho_{W}(\varphi, t)$
\be
 \rho_W(x,p,t) = \frac{1}{\pi\hbar}\int dy~\langle x-y|\hat{\rho}(t)|x +y \rangle e^{2ipy/\hbar},
\ee
which, although not strictly a density, reduces to it in the $\hbar\rightarrow 0$ limit.

Its evolution equation
 \be
\rho_W (\varphi_f, t_f) = \int {\cal D}\varphi_i~K_{qu}(\varphi_f,t_f|\varphi_i,t_i)\rho_W(\varphi_i,t_i)
\ee
is determined by the kernel $K_{qu}$, which we define on the same extended phase-space as
\be
 K_{qu}(\varphi_f,t_f|\varphi_i,t_i) = \int_{\varphi_i}^{\varphi_f} {\cal D}\varphi^a {\cal D}\Pi_a~\exp\bigg\{i S_{qu}[\varphi,\Pi]\bigg\},
 \label{qukern}
 \ee
with
\be
S_{qu}(\varphi,\Pi) =\int_{t_i}^{t_f}dt~{\cal L}_{qu}(\varphi,\Pi)=\int_{t_i}^{t_f}dt~({\dot\varphi^a}\Pi_b - {\cal H}_{qu}(\varphi,\Pi))
\ee

The quantum Hamiltonian ${\cal H}_{qu}(\phi,\Pi)$ is a Planckian finite-difference discretisation of ${\cal H}_{cl}(\phi,\Pi)$, to which it reduces as $\hbar\rightarrow 0$.
Several choices are possible. For the reasons given in \cite{gozzi2a,gozzi2b} we follow these authors in taking
\ba
 {\cal H}_{qu}(\varphi,\Pi) &=& -\frac{1}{2\hbar} [H_{cl}(\varphi^a +\hbar\omega^{ab}\Pi_b) - H_{cl}(\varphi^a -\hbar\omega^{ab}\Pi_b)].
\ea
We note that ${\cal H}_{qu}(\varphi,\Pi) = {\cal H}_{cl}(\varphi,\Pi)$ for quadratic (and linear and constant) potentials.

Although we appreciate that the paths themselves can be less important than the ways in which they are put together, what singles out complex classical mechanics (CCM) is the importance attached to individual paths in tracking their times of passage through the important regions of the complexified classical potential landscape, as in \cite{bender3}. {\it A priori}, we take the same stance here, in assuming that $K_{qu}$ is dominated by solutions to
\be
 \delta S_{qu}[\varphi,\Pi] = 0.
 \label{statph}
\ee
That is , we treat $S_{qu}[\varphi,\Pi]$ as a quasi-classical theory in its own right, which we shall term {\it mean-field quantum mechanics} (MFQM).
Since $S_{qu} = O(\hbar^0)$, (\ref{statph}) is a stationary phase approximation with no small parameter, and therefore to be taken circumspectly.

In what follows we compare MFQM to CCM. Although they do not match they have suggestive similarities. To cast $S_{qu}$ in a more familiar form, we reproduce Marinov \cite{marinov} by introducing new phase space variables:
\be
 \xi^a := \hbar\omega^{ab}\Pi_b.
\ee
$K_{qu}$ then takes the integral form
\be
 K_{qu}(\varphi_f,t_f|\varphi_i,t_i) = \int_{\phi_i}^{\phi_f} {\cal D}\varphi^a {\cal D}\xi^a~\exp\bigg\{\frac{i}{\hbar} S_{M}[\phi,\xi]\bigg\},
 \label{kqu2}
\ee
where
\be
S_{M}[\varphi,\xi] = \int_{t_i}^{t_f}dt~{\cal L}_{M}(\phi,\xi)
\ee
with
\be
 {\cal L}_{M}(\phi,\xi) = {\dot\varphi^a}\omega_{ab}\xi^b + \frac{1}{2}[H_{cl}(\varphi + \xi) - H_{cl}(\varphi - \xi)]
\ee
We stress again that the formalism of (\ref{kqu2}) is misleading, in that it suggests that the stationary phase approximation is also a small-$\hbar$ result, whereas $S_M$ is $O(\hbar )$. Despite that, there has been considerable work that successfully utilises the stationary-phase solutions [see \cite{dittrich} and applications cited therein].

\subsection{ Structure of MFQM}

Let us define
$ \xi^1:=y,~\xi^2 := q$
and
\begin{eqnarray}
 H^{\pm} &:=& \frac{1}{2}[H_{cl}(\varphi + \xi) \pm H_{cl}(\varphi - \xi)]
\end{eqnarray}
with $\varphi \pm \xi = (x\pm y,p\pm q)$.
We rearrange the original extended phase-space $(\varphi, \Pi)$ into the 4D phase space $X$:
\\
%\vspace{0.2cm}
 \be
    X^1:=x,~~~~~~ X^2:=y,~~~~~~X^3:=p,~~~~~~X^4=q.
    \label{w}
    \nonumber
    \ee
On $X$ we introduce the Poisson bracket  $\lpb\cdot,\cdot\rpb$:
\be
    \lpb A,B\rpb:= \Omega^{ab}~
    \partial_{a}A\:\partial_{b}B,
    \label{syp-st}
    \ee
where
    \be
    \Omega = \left(\begin{array}{cc}
    0 & I\\
    -I & 0 \end{array}\right).
    \label{omega}
    \ee
%    whence
% \be
% \lpb X^a, X^b\rpb = \Omega^{ab},
% \label{commX}
% \ee
Hamilton's equations, which reduce to $\delta S_{qu} = 0$, then take the form
\be
 {\dot{X}^a} = \lpb X^a, H^+(X)\rpb = \Omega^{ab}\partial_b H^+(X).
 \label{Hamilton}
 \ee

 Since $H^+$ and $H^-$ are related by
  $\partial_a H^- = \Gamma_a^b\partial_b H^+$,
where
  \be
    \Gamma =\left(\begin{array}{cc}
    \sigma_1 & 0\\
    0 & \sigma_1\end{array}\right)
    \label{Lambda2}
    \ee
it follows that
\be
   \lpb H^-, H^+\rpb = \Omega^{ac}\Gamma^b_c~
    \partial_{a}H^+\:\partial_{b}H^+ = 0,
   \label{constraint}
 \ee
 since $\Omega\Gamma$ is antisymmetric. $H^- = const$ is a first class constraint upon the effective classical theory.

 We note that there is an equivalence between $H^+$ and $H^-$ in that, with respect to a slightly different symplectic matrix $\Omega'$, we could equally derive the equations of motion from $H^-$ as
  \be
 {\dot{X}^a} = \lpb X^a, H^-(X)\rpb' = \Omega'^{ab}\partial_b H^-(X).
 \label{Hamilton2}
 \ee
 This latter is more in accord with the closed timepath formalism (CTP), defined in the extended $(x, y)$ coordinate space, for which $H^-$ is the relevant Hamiltonian prior to momentum integration. We shall not pursue this further.

 The classical limit is straightforward. Remember that
 $y = \hbar{\bar y},\,\,\,\,\,\,q = \hbar{\bar q}$
where ${\bar y},{\bar q}$ are the $O(\hbar^0)$ KvN conjugate variables to $p$ and $x$.
As $\hbar\rightarrow 0$ then $y,q\rightarrow 0$,
 as does
 \be
 H^- = O(\hbar)\rightarrow 0.
 \ee
 At the same time we get the contraction $H^+\rightarrow H_{cl},\,\,\,\Omega\rightarrow\omega$.

\section{ Complex Classical Mechanics (CCM)}

We now consider complex phase space, repeating the analysis of Smilga \cite{smilga}, taking
the complex extension of the real phase-space as
\be
x\rightarrow {\cal Z} = x + iy,\,\, p\rightarrow {\cal P} = p - iq.
 \ee

The Hamiltonian $H_{cl}$ is decomposed as
\be
H_{cl}(p,x)\rightarrow H_{cl}({\cal P}, {\cal Z}) = H_{R}({\cal P}, {\cal Z}) + i H_I({\cal P}, {\cal Z}),
\ee
where
\begin{eqnarray}
 H_R &=& \frac{1}{2}[H_{cl}(x+iy,p-iq) + H_{cl}(x-iy,p+iq)],
 \\
 H_I &=& \frac{1}{2i}[H_{cl}(x+iy,p-iq) - H_{cl}(x-iy,p+iq)].
\end{eqnarray}

To see the symplectic structure  of CCM we label the phase-space variables $X^a$ as before:
\be
    X^1:=x,~~~~~~ X^2:=y,~~~~~~X^3:=p,~~~~~~X^4=q.
    \label{w}
    \ee
    and introduce an identical Poisson bracket $\lpb\cdot,\cdot\rpb$ to (\ref{syp-st}):
\be
    \lpb A,B\rpb:= \Omega^{ab}~
    \partial_{a}A\:\partial_{b}B,
    \label{syp-st2}
    \ee
for the same $\Omega$.
Hamilton's equations are now \cite{smilga}
 \be
 {\dot{X}^a} = \lpb X^a, H_R\rpb = \Omega^{ab}\partial_bH_R(X).
 \label{Hamilton}
 \ee
$H_R$ and $H_I$ are related by Cauchy-Reimann as
\be
  \partial_a H_I = \Lambda_a^b\partial_b H_R,
  \label{cauchy}
 \ee
where
    \be
    \Lambda =\left(\begin{array}{cc}
    -i\sigma_2 & 0\\
    0 & i\sigma_2\end{array}\right).
    \label{Lambda2}
    \ee
It follows that
\be
   \lpb H_I, H_R\rpb = (\Omega\Lambda)^{ab}~
    \partial_{a}H_R\:\partial_{b}H_R = 0,
   \label{constraint}
 \ee
since $\Omega\Lambda$ is also antisymmetric.
That is, as before we have a constrained system with first-order constraint $H_I = constant$.

 \section{ CCM v. MFQM:}
 There are obvious similarities between the two approaches, a consequence of the identities
 \begin{eqnarray}
   H^+(x,iy,p,-iq) &=& H_R(x,y,p,q)
   \nonumber
   \\
    H^-(x,iy,p,-iq) &=& iH_I(x,y,p,q).
    \label{sim}
 \end{eqnarray}
 Since
 \be
 \lpb y, q\rpb = \lpb iy, -iq\rpb
\nonumber
 \ee
we can see why the symplectic structure remains unchanged.
\\
\\
The similarity between the two formalisms is very apparent for the SHO, with Hamiltonian
\be
 H_{cl} = \half(p^2 +x^2).
 \nonumber
\ee
In the $x-y$ plane we find identical solutions for $x$ and $y$ in {\it both} CCM and MFQM.
These are tilted ellipses
\be
x = A\sin (t + \alpha_1),\,\,\,\,y = B\sin (t + \alpha_2),
\nonumber
\ee
leading to identical $H^-$ and $H_I$,
\be
H^- = H_I = AB\cos (\alpha_1 -\alpha_2)
\nonumber
\ee
This suggests a possible role for $\hbar$ in CCM for this and other potentials.
We remember that  $y=\hbar{\bar y},\,q = \hbar{\bar q}$ in MFQM, which encourages us to take $y=\hbar{\bar y},\,q = \hbar{\bar q}$ in CCM. Then

 \begin{itemize}
 \item
  CCM can now distinguish between 'large' and 'small' systems. The conventional classical limit is simply understood as the recovery of real phase space.
 \item
  With $\Im{m E}$ now $O(\hbar)$ the empirical CCM tunnelling observation $\Im{m E}\, \Delta t \approx const.$ is understood as the quantum uncertainty relation $\Delta E\Delta t = O(\hbar)$.
 \item
  The CCM tunnelling results in \cite{bender3} are understood as anomalous behaviour in the  limit $\Im{m E}\rightarrow 0$. We now understand this as the familiar small $\hbar$ behaviour when looking for the persistence of 'quantum' effects.
 \end{itemize}

Looking at more general potentials,
even in \cite{marinov} and \cite{gozzi2a} it was appreciated that the extra dimensions in MFQM permitted tunnelling without instantons. However, this should not blind us to strong differences between the two formalisms, for which the factors of $i$ are crucial. Most importantly, the constant energy surfaces of MFQM are {\it bounded}, whereas those of CCM are {\it unbounded}. As a result, individual particle trajectories  in CCM go to infinite distances in the $x-y$ plane (and back) in finite time, whereas those of MFQM are always bounded.

To see the effects of this boundedness, it is
convenient (with the former) to work with $z_{\pm} = \varphi \pm \xi$, since $H_{cl}(z_{\pm})$ are individually conserved:
\be
 {\dot z}_{\pm}^a = \omega^{ab}\frac{\partial H_{cl}(z_{\pm})}{\partial z_{\pm}^b}
 \nonumber
 \ee
 We can think of the classical paths for $z_{\pm}$ as the tips of  chords of length $O(\hbar)$
  whose midpoints are quantum paths, only constrained by boundary conditions. In the Lagrangian formalism (on integrating out $p,q$) this is the familiar closed time-path approach.

 For example,  now consider 'tunnelling' in a double-well potential with binding energy $E_0$.
In CCM we take $H_R = E_n < E_0$ to match the energy of a bound state of the potential and fix $H_I = \Im{m E} = \Delta E \neq 0$. If, for example, we then consider paths with starting points $y=0,\,\,x=x_0$ the particle flips from well to well in a 'symmetric' way using the additional dimensions \cite{bender3}.
 If we now measure the ratio of the time the particle spends in each well as $\Delta E\rightarrow 0$ we can compare this with the QM results obtained from the wavefunctions as $H_I = \Im{m E} \rightarrow 0$. See \cite{bender3} for details. While not being compelling, the results are certainly interesting.

 However, the situation is very different in MFQM because of the boundedness of the constant energy surfaces. We now have
$$H^+ = {\half}[H_{cl}(z_+)+H_{cl}(z_-)].$$
As shown in \cite{dittrich}, to proceed we choose  $H^+ = E < E_0$ with $H_{cl}(z_+) > E_0$ and $H_{cl}(z_-) < E_0$ (or v.v) and fix $H^- = \Delta E \neq 0$. That is, one end of the chord is trapped in a well, while the other end is free.
The initial conditions mean that the particle flips from well to well 'asymmetrically' using additional dimensions. For this reason it is not sensible to attempt to measure the relative time the particle spends in each well as $\Delta E\rightarrow 0$. These trajectories make fundamentally clear a profound difference between CCM and MFQM that lies at the heart of quantum mechanics. For all the qualitative similarities between CCM and quantum mechanics, the defining ingredient of the latter is {\it interference}. This is immediately clear from the evolution equation (\ref{qukern}) for {\it real} density matrices, which demands that $K_{qu}$ be real. At the very least, if ($\varphi_{cl},\xi_{cl}$) are solutions to (\ref{Hamilton}) then so are ($\varphi_{cl},-\xi_{cl}$) and both solutions have to be combined ($z_+\leftrightarrow z_-$) in (\ref{qukern}), with the appropriate determinant of small fluctuations around the classical solutions. Unfortunately, this makes a direct comparison between the approaches impossible. Thus we can't use MFQM directly to criticise CCM, or vice-versa, as we had hoped.

  %It might be thought that there is a potential issue with interference in the Hamiltonian formulation of classical mechanics of Section 2 for which there are potentially more observables,  i.e. Hermitian functions of $\varphi$ and $\Pi$, than in the standard approach to (real) classical mechanics. Because of the non-commutativity of $\varphi$ and $\Pi$, as shown in (\ref{CCR}), interference looks to be possible, although we know this is not the case. In fact, not all these extra observables are invariant under a set of universal local symmetries which appear once the formalism is extended to differential forms on phase space\cite{nove} and, because of this, have to be removed. As Gozzi has shown, this makes the superposition of states in (real) CM impossible \cite{gozzi4}. Whether this is equally true for complex classical mechanics is unclear.

In fact, we need more than interference between quasi-classical paths in MFQM, as can be seen from the simple $V(x) = -{\half}x^2$ potential. Quasi-classical solutions in MFQM show that the particle rebounds for $E < 0$ i.e. there is no 'tunnelling' despite the extra dimension. Nonetheless, tunnelling happens in MFQM because of state preparation \cite{balazs}, whereby the tail of the wavefunction crosses the $x-p$ separatrix and is stretched to the other side of the potential hill.

We conclude with a comment on the first order constraints $H^- = 0$ and $H_I = 0$ that the formalisms possess. To accommodate them fully requires gauge fixing in the path integrals (or a change of bracket). At our simple level of comparison this is unnecessary, but see Smilga \cite{smilga} for a detailed discussion of gauge-fixing for CCM.

\section{Quantisation as dimensional reduction}

So far we have compared the classical trajectories of particles in complex phase space with the classical trajectories $z_{\pm}$ of the ends of the chords whose midpoints are the stationary phase solutions that define what we have called mean-field quantum mechanics. The doubling of the degrees of freedom by having to take account of two chord ends has its counterpart in the doubling of degrees of freedom by making phase space complex. Since the chords are of length $O(\hbar)$ we recover real classical mechanics from mean-field quantum mechanics trivially and, if the complex phase-space coordinates are, equally, $O(\hbar)$, recover real classical mechanics from complex classical mechanics at the same time. However, the real significance of the doubling of degrees of freedom in the chords is that these new coordinates are the quantum generalisations of the classical 'momenta' conjugate to the real phase space variables in the KvN Hilbert space formalism of classical mechanics.

This suggests an entirely different approach to 'deriving' quantum mechanics from classical mechanics due to Gozzi \cite{gozzi1a}, that we sketch below. It relies on the extension of the formalism of Section 2 to differential forms that we have already alluded to above.
For the moment we stay firmly with classical mechanics in real phase space. The rightmost integral of (\ref{cl1}) contains the Jacobian
 \be
 J := \textrm{det}(\delta_b^a\partial_t-\omega^{ac}\partial_c\partial_bH)
 \ee
which we have set to unity, as a consequence of the incompressibility of phase space. However, we can equally express it
 in terms of $2+2$
ghost-field Grassmann variables $(c^a,\bar{c}_a)$  as \cite{gozzi1a}
%%%
\ba
\displaystyle
J=\int {\cal D}\bar{c}_a{\cal D}c^a \exp\left[-\int dt
\bar{c}_a[\delta_b^a\partial_t
-\omega^{ac}\partial_c\partial_bH]c^b\right]. \label{II-7a}
\ea

%%%
Inserting this in the integrand of the rightmost path integral in (\ref{cl1}) enables us to write the kernel $K_{cl}$ as
%%%
\begin{equation}
K_{cl} = \int {\cal D}X^a {\cal D}\Pi_a{\cal
D}\bar{c}_a {\cal D}c^a \, \exp\left[i\int dt \, \bar{\cal
L}_{cl}\right] \label{II-8a}
\end{equation}
%%%
where we have dropped explicit mention of boundary conditions to simplify the notation. In (\ref{II-8a}) ${\cal L}_{cl}$ of (\ref{lcl}) is replaced by
%%%
\ba
\displaystyle \bar{\cal
L}_{cl}&=&\Pi_a[\dot{X}^a-\omega^{ab}\partial_bH]+
i\bar{c}_a[\delta_b^a\partial_t-\omega^{ac}\partial_c\partial_bH]c^b
\nonumber
\\
&=&\Pi_a\dot{X}^a +
i\bar{c}_a\dot{c}^a - \bar{\cal
H}_{cl},
\label{Lbar}
\ea
where the Hamiltonian associated with $\bar{\cal L}$ of (\ref{Lbar}) is now
%%%
\begin{equation}
\displaystyle \bar{\cal
H}_{cl}=\Pi_a\omega^{ab}\partial_bH+i\bar{c}_a\omega^{ac}\partial_c\partial_bHc^b.
\label{Hbar}
\end{equation}
The equations of motion that follow from $\widetilde{\cal
L}_{cl}$ show that the $(c^a,\bar{c}_a)$ are conjugate Jacobi variables in the sense of (\ref{CCR}) and that $\bar{\cal
H}_{cl}$ is the Lie derivative of Hamiltonian flow associated with $H_{cl}$ \cite{gozzi1a}.

  All that matters for this discussion is that, on introducing Grassmann partners $(\theta,{\bar\theta})$ to time $t$, Gozzi \cite{gozzi1a} constructed superspace phase space variables $\Phi = (X,P)$
\be
 \Phi^a(t,\theta,{\bar\theta}):= \varphi^a(t) + \theta c^a(t) + {\bar\theta}{\bar c}^a(t) + i{\bar\theta}\theta\omega^{ab}\Pi_b
\ee
We note that ${\bar\theta}\theta$ has the dimensions of inverse action. It follows \cite{gozzi1a} that
\be
i\int d\theta d{\bar\theta}~H_{cl}(\Phi) = {\cal H}_{cl}(\varphi, \Pi).
\ee
Similarly,
\be
i\int dt d\theta d{\bar\theta}~L_{cl}(\Phi) = \int dt~{\cal L}_{cl}(\varphi, \Pi),
\ee
where
\be
L_{cl} = p{\dot x} - H_{cl}.
\ee

The theory possesses BRS invariance. We retrieve {\it quantum mechanics} from {\it classical mechanics} by making  the dimensional reduction \cite{gozzi5}
\be
 i\hbar \int dt d\theta d{\bar\theta}\rightarrow \int dt
\ee
in superspace, together with $(X,P)\rightarrow (x,p)$ in phase space. [See also \cite{jizba} for the relationship of this approach to 't Hooft's derivation of quantum from classical physics \cite{tHa,tHb}.]

This analysis permits a natural extension to classical mechanics in complex phase space, with co-ordinates $X^a$ of (\ref{w}). The first step is to double this already doubled phase space by introducing four conjugate variables ${\bar\Pi}_a$, to be supplemented by (now $4+4$) Grassmann variables $(c^a,\bar{c}_a)$. We then look for BRS symmetry (and its breaking) in this extended space. If dimensional reduction is possible, the outcome will be quantum mechanics defined on complex phase space i.e. we shall be looking at probability densities in the complex plane \cite{bender4}. This is to be distinguished from the results of \cite{bender3}, in which comparison is made between CCM and quantum mechanics in the real plane. However, it will allow for a discussion of pseudo-Hermitian Hamiltonians, whose quantum mechanics has already been described in detail \cite{BB,B,B2} and with which much of the discussion of \cite{smilga} was concerned. Work on this is in preparation \cite{rivers3}.

 \section{ Conclusions:}

Our conclusions derived from these explorations of path integrals are somewhat schizophrenic. On the one hand, if we take the behaviour of particles in complex classical mechanics (CCM) as really reflecting attributes of quantum mechanics (QM), then formal similarities between CCM and mean-field quantum mechanics (MFQM) suggest that the complex dimensions of CCM should be taken as $O(\hbar)$. This resolves several issues with CCM, such as how classical mechanics distinguishes between classical and quantum systems, and how to interpret uncertainty relations.

On the other hand,  the differences between CCM and MFQM are at least as important as the similarities. In particular, the boundedness of the energy surfaces of the latter (in comparison to the unbounded nature of those of the former) mean that the trajectories are very different, even though each permits tunnelling without instantons and we know that, in many circumstances, MFQM gives a reliable description of quantum mechanical particles. However, for MFQM quantum superposition is a necessary ingredient, whereas (for real phase space at least) superposition plays no role in the path integrals of classical mechanics \cite{nove,gozzi4}.

Much of the original work on CCM was concerned with comparing its results to quantum mechanics in the {\it real} plane e.g. \cite{bender3}. As an alternative, we have raised the possibility of looking for supersymmetric realisations of CCM, with the potential of getting quantum mechanics in complex phase space. This will be pursued elsewhere \cite{rivers3}.

\section*{Acknowledgments}

We acknowledge stimulating discussions with Ennio Gozzi and thank him for access to unpublished notes on 'tunnelling without instantons' (December, 1995). We also thank Daniel Hook for empirically demonstrating the nature of 'tunnelling' paths in MFQM.

\end{document}